\documentclass[aps,prl,amsmath,amssymb,twocolumn]{revtex4-1}

\usepackage{graphicx}
\usepackage{dcolumn}
\usepackage{footmisc}

\usepackage{slashed}
\usepackage{bm}
\usepackage{hyperref}
\usepackage{xcolor}
\usepackage{float}
\begin{document}
\setlength{\abovedisplayskip}{5pt}
\setlength{\belowdisplayskip}{5pt}
\setlength{\abovedisplayshortskip}{5pt}
\setlength{\belowdisplayshortskip}{5pt}

\preprint{}

\title{Probing New Physics from Neutrinos at Dark Matter Direct Detection Experiments}

\author{Gonzalo Herrera}
\affiliation{Center for Neutrino Physics, Department of Physics, Virginia Tech, Blacksburg, VA 24061, USA}


\begin{abstract}
Dark matter direct detection experiments have become excellent low-energy neutrino detectors. We present a few novel ideas to probe Beyond the Standard Model physics from neutrinos at these experiments. First, we discuss signatures arising from the Migdal effect induced by solar neutrinos via Coherent Elastic Neutrino Nucleus Scatterings (CE$\nu$NS) at the detector. Second, we discuss that placing a radioactive source near a liquid xenon detector may allow to observe the anapole moment of neutrinos expected in the Standard Model via neutrino-electron scattering. Third, we point out that neutrinos coupled to a light dark sector charged under a dark $U(1)'$ symmetry could feature enhanced electromagnetic properties at the loop-level. We further discuss the testability of this scenario with direct detection experiments sensitive to solar neutrinos.\\

\href{https://indico.fnal.gov/event/63406/contributions/297406/}{Invited Contribution to the 25th International Workshop on Neutrinos from Accelerators (NuFact)}
\end{abstract}

\maketitle


\section{\label{sec:introduction} Introduction}
Low-energy neutrinos may hold surprises. When neutrinos scatter off nuclei or electrons in a detector with small momenta, the cross section may be enhanced with respect to the Standard Model (SM) expectation. This fact has been long known, the prominent example being a magnetic moment interaction of the neutrino. Similarly, new light mediators would also induce an enhanced interaction rate at low energies, due to the propagator suppression with the momentum transfer of the scattering process. Such interactions, although postulated decades ago, have yet not been observed at dedicated neutrino experiments.

Low-threshold and low-background dark matter direct detection experiments, originally designed to search for Weakly Interacting Massive Particles, have recently become sensitive to scatterings from solar neutrinos. Solar neutrino-electron scatterings have been included in background fits at direct detection experiments for a few years already, \textit{e.g} Refs. \cite{XENON:2022ltv, LZ:2022lsv, PandaX:2024cic}, accounting for tens to hundreds of events at the detector. Coherent Elastic Neutrino Nucleus scattering (CE$\nu$NS) recently received a dedicated analysis, with a significance over background of 2.6-2.7 $\sigma$, see Refs. \cite{PandaX:2024muv, XENON:2024ijk}.

The last decade has seen a variety of works aiming to constrain Beyond the Standard Model (BSM) properties of neutrinos with dark matter direct detection experiments, both via neutrino-electron scattering and via CE$\nu$NS at the detector. These efforts proved that dark matter direct detection experiments can provide stronger constraints on some BSM models of neutrinos than complementary probes.
However, past literature neglected the impact that new physics in the neutrino sector could have on the ionization signal arising from the Migdal effect. When neutrinos scatter off a nucleus, this can acquire a sufficiently large velocity to ionize electrons in the shell, resulting in an additional ionization signal at the detector following the nuclear recoil. In Ref. \cite{Herrera:2023xun}, we studied this effect, finding that BSM interactions of neutrinos can induce an ionization rate via the Migdal effect with a significantly different spectal shape than the SM contribution. Further, in Ref. \cite{Blanco-Mas:2024ale} we showed that the contribution from the Migdal effect yields a sizable number of events in the region of interest of direct detection experiments where CE$\nu$NS was observed, highlighting that these experiments are already sensitive to the yet unobserved Migdal effect.

The Migdal effect is strongly suppressed at recoil energies above a few keV, where neutrino-electron scatterings clearly dominates over CE$\nu$NS. In the second part of these Proceedings, focusing on the kinematic regime above $\sim 1$ keV, we discuss prospects to detect the anapole moment of neutrinos predicted in the SM via neutrino-electron scatterings at the detector from placing a radioactive ${}^{51}$Cr source near a liquid xenon detector. We base this discussion on Ref. \cite{Herrera:2024ysj}.

Finally, we discuss a simple model where the anapole and magnetic moment of neutrinos can be enhanced with respect to the SM expectation via a light millicharged dark sector. We show that direct detection experiments can provide world-leading laboratory constraints on such millicharged neutrinophilic light sectors, via the absence of signal at the detector from an enhanced magnetic or anapole moment.
\section{\label{sec:migdal} Migdal effect from neutrinos}
In nuclear recoils involving large energy transfers, the velocity acquired by the nucleus can be large enough for some electrons in the shells to be ionized. This was first pointed out by Arkady Beynusovich Migdal in 1939, in Ref. \cite{migdal:1939svj}. After a neutrino scatters off a nucleus, the ionization rate from the Migdal effect is given by (see Ref. \cite{Ibe:2017yqa} for details)
\begin{equation}
 \left |Z_{ion}(E_{er})  \right |^{2}=\frac{1}{2\pi}\sum_{n,l}\int dE_{er}\frac{d}{d E_{er}}p(nl \rightarrow E_{er}) 
\end{equation}
where the ionization probability of an electron in the orbital $(n,l)$ is denoted by $p$, and $E_{er}$ denotes the final ionized electron recoil energy. In order to find the total ionization rate arising from CE$\nu$NS, one needs to convolve the nuclear recoil rate with the electron ionization rate from the Migdal effect.
\begin{figure}[t!]
		\centering		\includegraphics[width=0.49\textwidth]{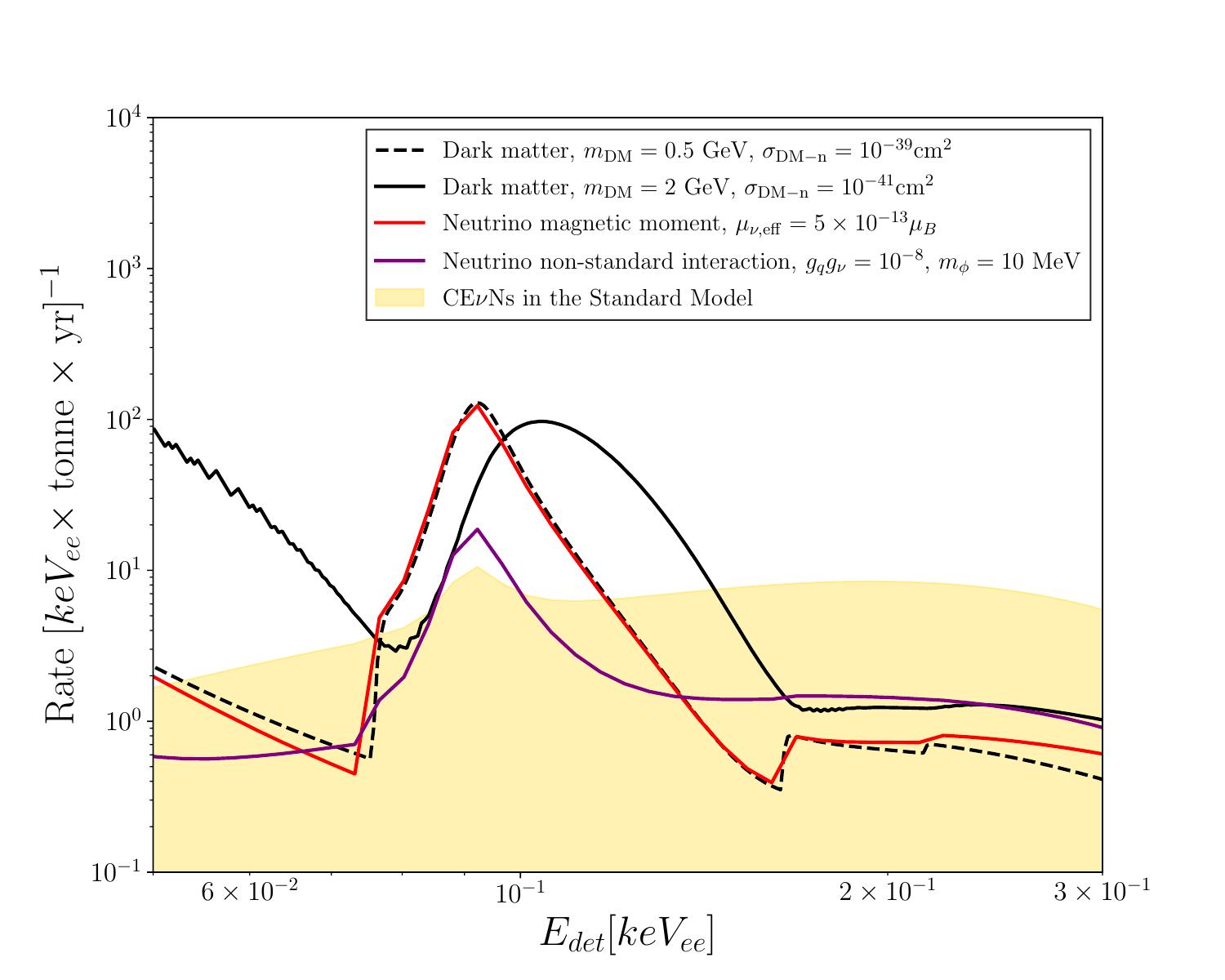}
		\centering
		\caption{Migdal ionization spectrum of xenon from CE$\nu$NS in the SM (yellow), via a large neutrino magnetic moment (red), a new light scalar mediator (purple) and from galactic dark matter (black). The ionization spectrum in the SM is roughly constant around 0.1 keV$_{ee}$. However, in the presence of new physics from neutrinos or light dark matter, the peak can be significantly enhanced w.r.t to the SM.}
	\label{fig:migdal}
\end{figure}
In Ref. \cite{Herrera:2023xun}, we calculated the Migdal ionization rate from solar neutrinos at a liquid xenon detector. Interestingly, we found that such ionization rate at the detector (in electron equivalent energy $E_{\rm det}$ [keV$_{ee}$]) is comparable to the ionization rate from neutrino electron scattering and from CE$\nu$NS in the range $E_{\rm det} \simeq 0.3-0.6$ keV$_{ee}$, with a distinct spectral shape compared to other contributions. Since dark matter direct detection experiments are already sensitive to events from neutrino-electron scattering and CE$\nu$NS around that energy range, this poses the question of whether the signal induced by the Migdal effect from neutrinos could also be distinguised. We demonstrated in Ref. \cite{Herrera:2023xun} that a significant number of events would be obtained with an exposure of $\sim 5$ tonne $\times$ year for an experiment with the characteristics of the XENONnT experiment, if the energy threshold for a combined scintillation-ionization (S1+S2) analysis were as low as 0.1 keV$_{ee}$. Furthermore, in Ref. \cite{Blanco-Mas:2024ale} we showed that current datasets from XENONnT and PANDAX-4T presenting a hint from CE$\nu$NS are already sensitive to $O(1-10)$ events arising from the Migdal effect. However, the Migdal-induced rate amounts to $\sim 10\%$ of the rate induced by CE$\nu$NS, such that the significance of such a signal is likely too small to be able to claim any hint. Furthermore, since the Migdal effect has not yet been observed from nuclear recoils of other SM particles, like neutrons, the validity of current theoretical calculations is yet to be assessed, since these calculations assume ionization probabilites for isolated atoms and certain values for the quenching factor at low energies, both of which present uncertainties in liquid xenon, see Refs. \cite{Bell:2021ihi,Xu:2023wev, bang2023migdal}.
However, the ionization signal arising from the Migdal effect may be enhanced in presence of new physics. In Ref. \cite{Herrera:2023xun} we calculated for the first time the Migdal ionization rate induced by neutrino electromagnetic properties and by new light mediators coupled to neutrinos and quarks. We found that in some instances the ionization rate can mask the rate induced by light dark matter particles from the galactic halo at the detector, but could be very different to the expected Migdal rate from neutrinos via weak interactions only. Crucially, new physics can induce a distinct peak around $E_{\rm det} \simeq 0.1$ keV$_{ee}$ that is not present in the SM, see Fig. \ref{fig:migdal}. Such a peak can arise from the ionization of electrons in the $n=4$ shell via $pp$ solar neutrinos, which is enhanced at small momentum transfer.

\section{\label{sec:anapole} A ${}^{51}$Cr source to detect the anapole moment of neutrinos}
Although neutrinos are electrically neutral, the Standard Model predicts electromagnetic interactions of neutrinos at the loop-level (see Ref. \cite{Giunti:2014ixa} for an useful review). One of the allowed interactions is given by the parity violating term $\sim \left(q^2 \gamma^\mu-q^\mu \slashed{q} \right) \gamma_5$ in the electromagnetic vertex of a photon and a neutrino. In the limit of small momentum transfer, this term leads to a neutrino anapole moment, first introduced by Zeldovich in 1958, see Ref. \cite{1958JETP....6.1184Z}. For ultrarelativistic neutrinos ($\gamma_{5} \rightarrow -1$), the anapole moment is equivalent to the charge radius of the neutrino, as they are related via
$a_{ee}=\left\langle r_{ee}^2\right\rangle/6$. The anapole moment of neutrinos can be computed in the SM, yielding a value for electron neutrinos of 
\begin{equation}
a_{ee}^{\mathrm{SM}} \simeq \frac{G_F}{24 \sqrt{2} \pi^2}\left(3-2 \log \frac{m_{e}^2}{m_W^2}\right) \simeq 6.8 \times 10^{-34} \mathrm{~cm}^2.
\end{equation}
Although this value may seem small, it is not suppressed by the neutrino mass, unlike the neutrino magnetic moment. The main drawback for a detection is its degeneracy with the uncertainty on the weak mixing angle, since the anapole moment induces a cross section with the same energy dependence as that of weak interactions, but modifying the weak mixing angle via
\begin{equation}
\sin ^2 \theta_W \rightarrow \sin ^2 \theta_W\left(1-2m_W^2a_{ee}\right).
\end{equation}
It turns out that reactor neutrino experiments and solar neutrino experiments are not sensitive yet to the SM value of the neutrino anapole moment, but lie remarkably close, placing limits one order of magnitude away from detection.
\begin{figure}[t!]
		\centering
        \includegraphics[width=0.49\textwidth]{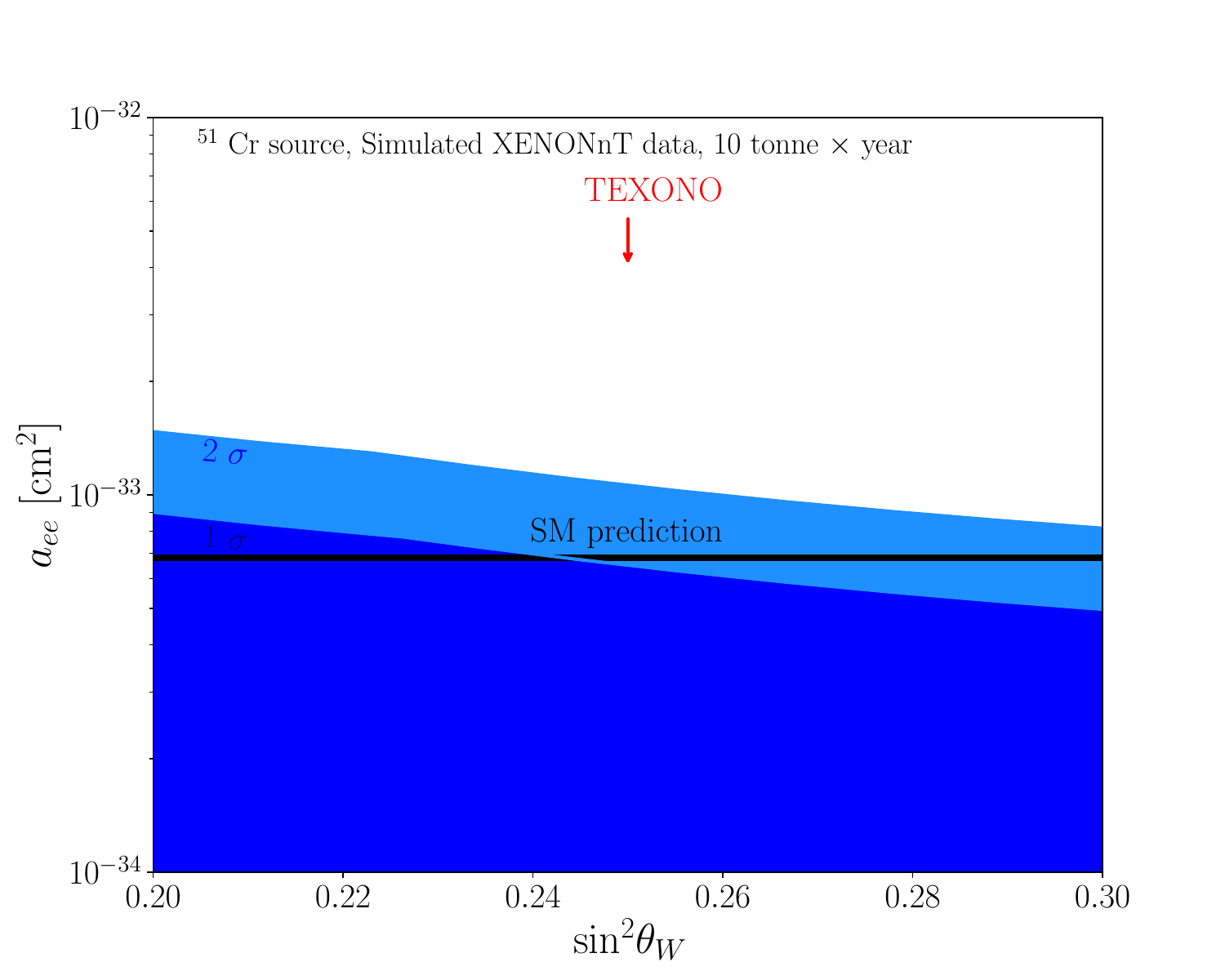}
		\centering
		\caption{Sensitivity contours on the electron neutrino anapole moment with XENONnT and a ${}^{51}$Cr source and a exposure of 10 tonne $\times$ year. The strongest complementary limit, from TEXONO, is shown for comparison, see Ref. \cite{TEXONO:2009knm}. We show the SM prediction on the electron neutrino anapole moment for comparison.}
	\label{fig:anapole}
\end{figure}
In Ref. \cite{Herrera:2024ysj}, we proposed placing a radioactive source near a large volume liquid xenon detector, to maximize the neutrino flux and enhance the prospects for a detection of the neutrino anapole moment. We focused our discussion on a ${}^{51}$Cr source, which had been previously used at neutrino experiments like GALLEX. A typical ${ }^{51} \mathrm{Cr}$ source produces electron neutrinos via the reaction ${ }^{51} \mathrm{Cr}+e^{-} \rightarrow{ }^{51} \mathrm{~V}+\nu_e$ with a half-life of 27.7 days, and the neutrino flux can overcome that from solar neutrinos when the source is placed close enough to the detector.

We demonstrated in Ref. \cite{Herrera:2024ysj} that 1$\sigma$ to $2\sigma$ sensitivity to the electron neutrino anapole moment can be achieved with a ${}^{51}$Cr source placed at 1m from the XENONnT detector, for an exposure of 10 tonne $\times$ year. This conclusion is robust against uncertainties on the weak mixing angle at low energies, see Fig \ref{fig:anapole}. Such an exposure could be achieved at the current XENONnT experiment by running the ${}^{51}$Cr source over a handful of cycles. For detector masses as those expected for the future XLZD experiment \cite{XLZD:2024nsu}, only one or two cycles of the Chromium source would be enough to reach such sensitivity.

\section{\label{sec:darkmoment} Dark neutrino moments from light loops}

The SM value of the neutrino magnetic moment is currently undetectable, due to its suppression by the neutrino mass, yielding a value for electron  (Dirac) neutrinos of
\begin{equation}
\mu_{ee} \simeq \frac{3 e G_{\mathrm{F}} m_{\nu_{e}}}{8 \sqrt{2} \pi^2} \simeq 3.2 \times 10^{-19}\left(\frac{m_{\nu_{e}}}{ \mathrm{~eV}}\right) \mu_{\mathrm{B}}
\end{equation}
where $\mu_{B}$ denotes the Bohr magneton. A similar value can be found for Majorana neutrinos. Due to the smallness of the (effective) electron neutrino mass, this value turns out really small, orders of magnitude away from current constraints.

Similarly, we discussed in the previous section that the expected value of the electron neutrino anapole moment is still undetectable due to its degeneracy with the uncertainty on the weak mixing angle. Thus, it is worth wondering whether there are Beyond the Standard Model scenarios where the value of the neutrino electromagnetic moments can be enhanced w.r.t the SM expectation. We propose in Refs. \cite{Herrera:2024ysj,Herrera_Shoemaker_2024} a simple model where this can be achieved. We consider a light (keV to GeV scale) sector where millicharged particles (say scalars $\phi$, vectors $V$ and fermions $\chi$) also couple to neutrinos, generating a dark neutrino moment at the loop level. This generates an effective electromagnetic moment via the kinetic mixing $\epsilon$ of the dark photon with the SM photon.
\begin{figure}[t!]
		\centering
		\includegraphics[width=0.49\textwidth]{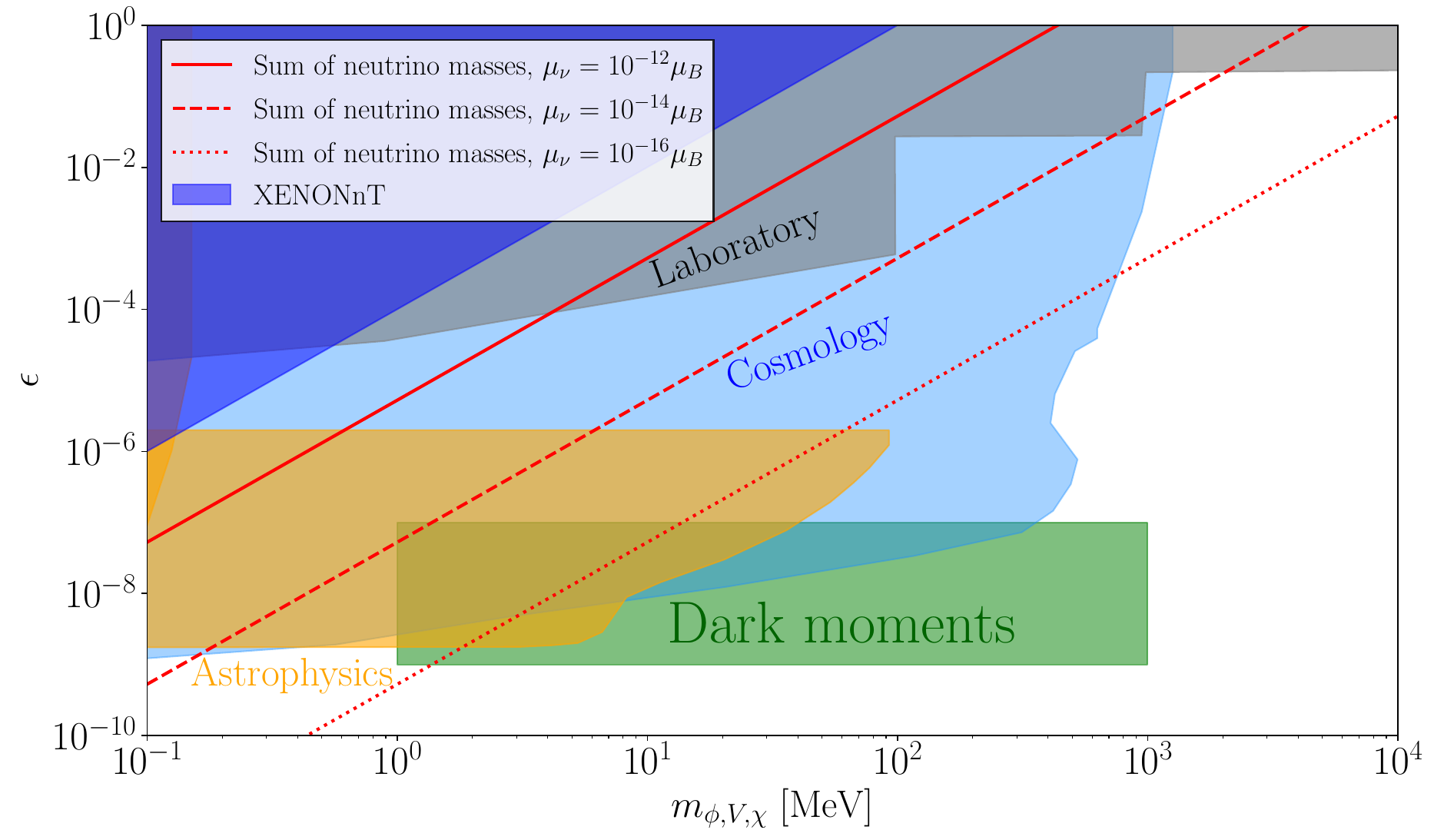}
		\centering
		\caption{Constraints on millicharged light particles from a variety of astrophysical, cosmological
and laboratory probes. For comparison, we show constraints in red on the off-diagonal neutrino magnetic moment generated by these light millicharged particles, from the requirement that
the sum of neutrino masses shall not exceed the cosmological bound $\sum m_{\nu}\lesssim 0.2$ eV. Further, the dark blue region is excluded by the XENONnT experiment. For comparison, we show in the dark green region the parameter space of dark moments that can yield enhanced neutrino magnetic moments.}
	\label{fig:dark}
\end{figure}
Although light millicharged particles are strongly constrained by a combination of laboratory, cosmological and astrophysical probes, we find untested regions of parameter space where this model can yield values of the neutrino anapole and magnetic moments that are sizably larger than in the SM. A tentative region where such \textit{dark neutrino moments} are possible is shown in Fig. \ref{fig:dark}. Constraints from the sum of neutrino masses are also shown, since the following relation is expected between kinetic mixing, magnetic moment and neutrino masses, see Ref. \cite{Lindner:2017uvt}
\begin{equation}
\frac{\sum m_\nu}{0.1 \mathrm{eV}} \sim \frac{1}{\epsilon}\left(\frac{\mu_\nu}{10^{-13} \mu_B}\right)\left(\frac{m_{\phi, V, \chi}}{\mathrm{GeV}}\right)^{-2}.
\end{equation}
Furthermore, we show in the Figure novel constraints obtained in Ref.\cite{Herrera_Shoemaker_2024} from the XENONnT experiment in this parameter space. These constraints are the strongest laboratory constraints on millicharged neutrinophilic particles with masses below $m_{\phi,V,\chi} \lesssim 1$ MeV.
\section{Conclusions}
We are entering an era where dark matter direct detection experiments are becoming world-leading low-energy neutrino detectors. Interactions of low-energy neutrinos at dark matter detectors may shed light on new physics arising from neutrino electromagnetic moments or light mediators. Here we have first discussed the prospects to detect Beyond the Standard Model interactions of neutrinos via the yet-unobserved Migdal Effect, demonstrating that this novel neutrino channel can provide distinct spectral features at experiments, see Ref. \cite{Herrera:2023xun}. 

Secondly, we have discussed that the anapole moment of the electron neutrino, as predicted in the Standard Model, may be observed at the 1-2$\sigma$ level if placing a radioactive ${}^{51}$Cr source near a large volume liquid xenon detector as that proposed for the future XLZD experiment. The time exposure required will be fairly small, of about two source lifetimes, see Ref. \cite{Herrera:2024ysj}.

Thirdly, we have discussed a model where the magnetic and anapole moments of neutrinos is enhanced w.r.t the SM, via a light millicharged sector that also couples to neutrinos. In this manner, we point out that direct detection experiments setting constraints on neutrino magnetic and anapole moments are also indirectly constraining millicharged neutrinophilic sectors in a meaningful manner, complementary to other laboratory probes, cosmology and astrophysics, see Refs. \cite{Herrera:2024ysj,Herrera_Shoemaker_2024}.

\section*{Acknowledgments}
We thank the organizers of \href{https://indico.fnal.gov/event/63406/}{NuFact 2024} for the successful conference. We thank the conveners of Working Group 5, Koun Choi, Julia Harz and Matheus Hostert for the invitation. We are especially grateful to Patrick Huber and Ian M. Shoemaker for collaboration and encouragement. We are also grateful to Pablo Blanco-Mas, Pilar Coloma, Joachim Kopp and Zahra Tabrizi for collaboration and useful discussions on the Migdal effect and light mediators. This work is supported by the U.S. Department of Energy Office of Science under award number DE-SC0020262.

\bibliography{references}

\begin{thebibliography}{19}%
\makeatletter
\providecommand \@ifxundefined [1]{%
 \@ifx{#1\undefined}
}%
\providecommand \@ifnum [1]{%
 \ifnum #1\expandafter \@firstoftwo
 \else \expandafter \@secondoftwo
 \fi
}%
\providecommand \@ifx [1]{%
 \ifx #1\expandafter \@firstoftwo
 \else \expandafter \@secondoftwo
 \fi
}%
\providecommand \natexlab [1]{#1}%
\providecommand \enquote  [1]{``#1''}%
\providecommand \bibnamefont  [1]{#1}%
\providecommand \bibfnamefont [1]{#1}%
\providecommand \citenamefont [1]{#1}%
\providecommand \href@noop [0]{\@secondoftwo}%
\providecommand \href [0]{\begingroup \@sanitize@url \@href}%
\providecommand \@href[1]{\@@startlink{#1}\@@href}%
\providecommand \@@href[1]{\endgroup#1\@@endlink}%
\providecommand \@sanitize@url [0]{\catcode `\\12\catcode `\$12\catcode `\&12\catcode `\#12\catcode `\^12\catcode `\_12\catcode `\%12\relax}%
\providecommand \@@startlink[1]{}%
\providecommand \@@endlink[0]{}%
\providecommand \url  [0]{\begingroup\@sanitize@url \@url }%
\providecommand \@url [1]{\endgroup\@href {#1}{\urlprefix }}%
\providecommand \urlprefix  [0]{URL }%
\providecommand \Eprint [0]{\href }%
\providecommand \doibase [0]{http://dx.doi.org/}%
\providecommand \selectlanguage [0]{\@gobble}%
\providecommand \bibinfo  [0]{\@secondoftwo}%
\providecommand \bibfield  [0]{\@secondoftwo}%
\providecommand \translation [1]{[#1]}%
\providecommand \BibitemOpen [0]{}%
\providecommand \bibitemStop [0]{}%
\providecommand \bibitemNoStop [0]{.\EOS\space}%
\providecommand \EOS [0]{\spacefactor3000\relax}%
\providecommand \BibitemShut  [1]{\csname bibitem#1\endcsname}%
\let\auto@bib@innerbib\@empty
\bibitem [{\citenamefont {Aprile}\ \emph {et~al.}(2022)\citenamefont {Aprile} \emph {et~al.}}]{XENON:2022ltv}%
  \BibitemOpen
  \bibfield  {author} {\bibinfo {author} {\bibfnamefont {E.}~\bibnamefont {Aprile}} \emph {et~al.} (\bibinfo {collaboration} {XENON}),\ }\href {\doibase 10.1103/PhysRevLett.129.161805} {\bibfield  {journal} {\bibinfo  {journal} {Phys. Rev. Lett.}\ }\textbf {\bibinfo {volume} {129}},\ \bibinfo {pages} {161805} (\bibinfo {year} {2022})},\ \Eprint {http://arxiv.org/abs/2207.11330} {arXiv:2207.11330 [hep-ex]} \BibitemShut {NoStop}%
\bibitem [{\citenamefont {Aalbers}\ \emph {et~al.}(2023)\citenamefont {Aalbers} \emph {et~al.}}]{LZ:2022lsv}%
  \BibitemOpen
  \bibfield  {author} {\bibinfo {author} {\bibfnamefont {J.}~\bibnamefont {Aalbers}} \emph {et~al.} (\bibinfo {collaboration} {LZ}),\ }\href {\doibase 10.1103/PhysRevLett.131.041002} {\bibfield  {journal} {\bibinfo  {journal} {Phys. Rev. Lett.}\ }\textbf {\bibinfo {volume} {131}},\ \bibinfo {pages} {041002} (\bibinfo {year} {2023})},\ \Eprint {http://arxiv.org/abs/2207.03764} {arXiv:2207.03764 [hep-ex]} \BibitemShut {NoStop}%
\bibitem [{\citenamefont {Zeng}\ \emph {et~al.}(2024)\citenamefont {Zeng} \emph {et~al.}}]{PandaX:2024cic}%
  \BibitemOpen
  \bibfield  {author} {\bibinfo {author} {\bibfnamefont {X.}~\bibnamefont {Zeng}} \emph {et~al.} (\bibinfo {collaboration} {PandaX}),\ }\href@noop {} {\  (\bibinfo {year} {2024})},\ \Eprint {http://arxiv.org/abs/2408.07641} {arXiv:2408.07641 [hep-ex]} \BibitemShut {NoStop}%
\bibitem [{\citenamefont {Bo}\ \emph {et~al.}(2024)\citenamefont {Bo} \emph {et~al.}}]{PandaX:2024muv}%
  \BibitemOpen
  \bibfield  {author} {\bibinfo {author} {\bibfnamefont {Z.}~\bibnamefont {Bo}} \emph {et~al.} (\bibinfo {collaboration} {PandaX}),\ }\href {\doibase 10.1103/PhysRevLett.133.191001} {\bibfield  {journal} {\bibinfo  {journal} {Phys. Rev. Lett.}\ }\textbf {\bibinfo {volume} {133}},\ \bibinfo {pages} {191001} (\bibinfo {year} {2024})},\ \Eprint {http://arxiv.org/abs/2407.10892} {arXiv:2407.10892 [hep-ex]} \BibitemShut {NoStop}%
\bibitem [{\citenamefont {Aprile}\ \emph {et~al.}(2024)\citenamefont {Aprile} \emph {et~al.}}]{XENON:2024ijk}%
  \BibitemOpen
  \bibfield  {author} {\bibinfo {author} {\bibfnamefont {E.}~\bibnamefont {Aprile}} \emph {et~al.} (\bibinfo {collaboration} {XENON}),\ }\href {\doibase 10.1103/PhysRevLett.133.191002} {\bibfield  {journal} {\bibinfo  {journal} {Phys. Rev. Lett.}\ }\textbf {\bibinfo {volume} {133}},\ \bibinfo {pages} {191002} (\bibinfo {year} {2024})},\ \Eprint {http://arxiv.org/abs/2408.02877} {arXiv:2408.02877 [nucl-ex]} \BibitemShut {NoStop}%
\bibitem [{\citenamefont {Herrera}(2024)}]{Herrera:2023xun}%
  \BibitemOpen
  \bibfield  {author} {\bibinfo {author} {\bibfnamefont {G.}~\bibnamefont {Herrera}},\ }\href {\doibase 10.1007/JHEP05(2024)288} {\bibfield  {journal} {\bibinfo  {journal} {JHEP}\ }\textbf {\bibinfo {volume} {05}},\ \bibinfo {pages} {288} (\bibinfo {year} {2024})},\ \Eprint {http://arxiv.org/abs/2311.17719} {arXiv:2311.17719 [hep-ph]} \BibitemShut {NoStop}%
\bibitem [{\citenamefont {Blanco-Mas}\ \emph {et~al.}(2024)\citenamefont {Blanco-Mas}, \citenamefont {Coloma}, \citenamefont {Herrera}, \citenamefont {Huber}, \citenamefont {Kopp}, \citenamefont {Shoemaker},\ and\ \citenamefont {Tabrizi}}]{Blanco-Mas:2024ale}%
  \BibitemOpen
  \bibfield  {author} {\bibinfo {author} {\bibfnamefont {P.}~\bibnamefont {Blanco-Mas}}, \bibinfo {author} {\bibfnamefont {P.}~\bibnamefont {Coloma}}, \bibinfo {author} {\bibfnamefont {G.}~\bibnamefont {Herrera}}, \bibinfo {author} {\bibfnamefont {P.}~\bibnamefont {Huber}}, \bibinfo {author} {\bibfnamefont {J.}~\bibnamefont {Kopp}}, \bibinfo {author} {\bibfnamefont {I.~M.}\ \bibnamefont {Shoemaker}}, \ and\ \bibinfo {author} {\bibfnamefont {Z.}~\bibnamefont {Tabrizi}},\ }\href@noop {} {\  (\bibinfo {year} {2024})},\ \Eprint {http://arxiv.org/abs/2411.14206} {arXiv:2411.14206 [hep-ph]} \BibitemShut {NoStop}%
\bibitem [{\citenamefont {Herrera}\ and\ \citenamefont {Huber}(2024)}]{Herrera:2024ysj}%
  \BibitemOpen
  \bibfield  {author} {\bibinfo {author} {\bibfnamefont {G.}~\bibnamefont {Herrera}}\ and\ \bibinfo {author} {\bibfnamefont {P.}~\bibnamefont {Huber}},\ }\href@noop {} {\  (\bibinfo {year} {2024})},\ \Eprint {http://arxiv.org/abs/2408.11904} {arXiv:2408.11904 [hep-ph]} \BibitemShut {NoStop}%
\bibitem [{\citenamefont {Migdal}(1939)}]{migdal:1939svj}%
  \BibitemOpen
  \bibfield  {author} {\bibinfo {author} {\bibfnamefont {A.}~\bibnamefont {Migdal}},\ }\href@noop {} {\bibfield  {journal} {\bibinfo  {journal} {Sov. Phys. JETP}\ }\textbf {\bibinfo {volume} {9}},\ \bibinfo {pages} {1163} (\bibinfo {year} {1939})}\BibitemShut {NoStop}%
\bibitem [{\citenamefont {Ibe}\ \emph {et~al.}(2018)\citenamefont {Ibe}, \citenamefont {Nakano}, \citenamefont {Shoji},\ and\ \citenamefont {Suzuki}}]{Ibe:2017yqa}%
  \BibitemOpen
  \bibfield  {author} {\bibinfo {author} {\bibfnamefont {M.}~\bibnamefont {Ibe}}, \bibinfo {author} {\bibfnamefont {W.}~\bibnamefont {Nakano}}, \bibinfo {author} {\bibfnamefont {Y.}~\bibnamefont {Shoji}}, \ and\ \bibinfo {author} {\bibfnamefont {K.}~\bibnamefont {Suzuki}},\ }\href {\doibase 10.1007/JHEP03(2018)194} {\bibfield  {journal} {\bibinfo  {journal} {JHEP}\ }\textbf {\bibinfo {volume} {03}},\ \bibinfo {pages} {194} (\bibinfo {year} {2018})},\ \Eprint {http://arxiv.org/abs/1707.07258} {arXiv:1707.07258 [hep-ph]} \BibitemShut {NoStop}%
\bibitem [{\citenamefont {Bell}\ \emph {et~al.}(2022)\citenamefont {Bell}, \citenamefont {Dent}, \citenamefont {Lang}, \citenamefont {Newstead},\ and\ \citenamefont {Ritter}}]{Bell:2021ihi}%
  \BibitemOpen
  \bibfield  {author} {\bibinfo {author} {\bibfnamefont {N.~F.}\ \bibnamefont {Bell}}, \bibinfo {author} {\bibfnamefont {J.~B.}\ \bibnamefont {Dent}}, \bibinfo {author} {\bibfnamefont {R.~F.}\ \bibnamefont {Lang}}, \bibinfo {author} {\bibfnamefont {J.~L.}\ \bibnamefont {Newstead}}, \ and\ \bibinfo {author} {\bibfnamefont {A.~C.}\ \bibnamefont {Ritter}},\ }\href {\doibase 10.1103/PhysRevD.105.096015} {\bibfield  {journal} {\bibinfo  {journal} {Phys. Rev. D}\ }\textbf {\bibinfo {volume} {105}},\ \bibinfo {pages} {096015} (\bibinfo {year} {2022})},\ \Eprint {http://arxiv.org/abs/2112.08514} {arXiv:2112.08514 [hep-ph]} \BibitemShut {NoStop}%
\bibitem [{\citenamefont {Xu}\ \emph {et~al.}(2024)\citenamefont {Xu} \emph {et~al.}}]{Xu:2023wev}%
  \BibitemOpen
  \bibfield  {author} {\bibinfo {author} {\bibfnamefont {J.}~\bibnamefont {Xu}} \emph {et~al.},\ }\href {\doibase 10.1103/PhysRevD.109.L051101} {\bibfield  {journal} {\bibinfo  {journal} {Phys. Rev. D}\ }\textbf {\bibinfo {volume} {109}},\ \bibinfo {pages} {L051101} (\bibinfo {year} {2024})},\ \Eprint {http://arxiv.org/abs/2307.12952} {arXiv:2307.12952 [hep-ex]} \BibitemShut {NoStop}%
\bibitem [{\citenamefont {Bang}(2023)}]{bang2023migdal}%
  \BibitemOpen
  \bibfield  {author} {\bibinfo {author} {\bibfnamefont {J.}~\bibnamefont {Bang}},\ }in\ \href {https://indico.cern.ch/event/1188759/contributions/5222299/} {\emph {\bibinfo {booktitle} {Talk at the UCLA Dark Matter 2023 Conference}}}\ (\bibinfo {address} {Los Angeles, CA},\ \bibinfo {year} {2023})\BibitemShut {NoStop}%
\bibitem [{\citenamefont {Giunti}\ and\ \citenamefont {Studenikin}(2015)}]{Giunti:2014ixa}%
  \BibitemOpen
  \bibfield  {author} {\bibinfo {author} {\bibfnamefont {C.}~\bibnamefont {Giunti}}\ and\ \bibinfo {author} {\bibfnamefont {A.}~\bibnamefont {Studenikin}},\ }\href {\doibase 10.1103/RevModPhys.87.531} {\bibfield  {journal} {\bibinfo  {journal} {Rev. Mod. Phys.}\ }\textbf {\bibinfo {volume} {87}},\ \bibinfo {pages} {531} (\bibinfo {year} {2015})},\ \Eprint {http://arxiv.org/abs/1403.6344} {arXiv:1403.6344 [hep-ph]} \BibitemShut {NoStop}%
\bibitem [{\citenamefont {{Zel'Dovich}}(1958)}]{1958JETP....6.1184Z}%
  \BibitemOpen
  \bibfield  {author} {\bibinfo {author} {\bibfnamefont {I.~B.}\ \bibnamefont {{Zel'Dovich}}},\ }\href@noop {} {\bibfield  {journal} {\bibinfo  {journal} {Soviet Journal of Experimental and Theoretical Physics}\ }\textbf {\bibinfo {volume} {6}},\ \bibinfo {pages} {1184} (\bibinfo {year} {1958})}\BibitemShut {NoStop}%
\bibitem [{\citenamefont {Deniz}\ \emph {et~al.}(2010)\citenamefont {Deniz} \emph {et~al.}}]{TEXONO:2009knm}%
  \BibitemOpen
  \bibfield  {author} {\bibinfo {author} {\bibfnamefont {M.}~\bibnamefont {Deniz}} \emph {et~al.} (\bibinfo {collaboration} {TEXONO}),\ }\href {\doibase 10.1103/PhysRevD.81.072001} {\bibfield  {journal} {\bibinfo  {journal} {Phys. Rev. D}\ }\textbf {\bibinfo {volume} {81}},\ \bibinfo {pages} {072001} (\bibinfo {year} {2010})},\ \Eprint {http://arxiv.org/abs/0911.1597} {arXiv:0911.1597 [hep-ex]} \BibitemShut {NoStop}%
\bibitem [{\citenamefont {Aalbers}\ \emph {et~al.}(2024)\citenamefont {Aalbers} \emph {et~al.}}]{XLZD:2024nsu}%
  \BibitemOpen
  \bibfield  {author} {\bibinfo {author} {\bibfnamefont {J.}~\bibnamefont {Aalbers}} \emph {et~al.} (\bibinfo {collaboration} {XLZD}),\ }\href@noop {} {\  (\bibinfo {year} {2024})},\ \Eprint {http://arxiv.org/abs/2410.17137} {arXiv:2410.17137 [hep-ex]} \BibitemShut {NoStop}%
\bibitem [{\citenamefont {Herrera}\ and\ \citenamefont {M.~Shoemaker}()}]{Herrera_Shoemaker_2024}%
  \BibitemOpen
  \bibfield  {author} {\bibinfo {author} {\bibfnamefont {G.}~\bibnamefont {Herrera}}\ and\ \bibinfo {author} {\bibfnamefont {I.}~\bibnamefont {M.~Shoemaker}},\ }\href@noop {} {\ }\Eprint {http://arxiv.org/abs/In progress} {In progress} \BibitemShut {NoStop}%
\bibitem [{\citenamefont {Lindner}\ \emph {et~al.}(2017)\citenamefont {Lindner}, \citenamefont {Radov\v{c}i\'c},\ and\ \citenamefont {Welter}}]{Lindner:2017uvt}%
  \BibitemOpen
  \bibfield  {author} {\bibinfo {author} {\bibfnamefont {M.}~\bibnamefont {Lindner}}, \bibinfo {author} {\bibfnamefont {B.}~\bibnamefont {Radov\v{c}i\'c}}, \ and\ \bibinfo {author} {\bibfnamefont {J.}~\bibnamefont {Welter}},\ }\href {\doibase 10.1007/JHEP07(2017)139} {\bibfield  {journal} {\bibinfo  {journal} {JHEP}\ }\textbf {\bibinfo {volume} {07}},\ \bibinfo {pages} {139} (\bibinfo {year} {2017})},\ \Eprint {http://arxiv.org/abs/1706.02555} {arXiv:1706.02555 [hep-ph]} \BibitemShut {NoStop}%
\end{thebibliography}%

\end{document}